\documentclass[aps,prb,twocolumn,showpacs]{revtex4}

\usepackage{graphicx}
\usepackage{amsmath}
\usepackage{amssymb}

\begin{document}

\title{Doping dependence of thermodynamic properties in cuprate superconductors}

\author{Huaisong Zhao, L\"ulin Kuang, and Shiping Feng}
\affiliation{Department of Physics, Beijing Normal University,
Beijing 100875, China}

\begin{abstract}
The doping and temperature dependence of the thermodynamic properties in cuprate superconductors is studied based on the
kinetic energy driven superconducting mechanism. By considering the interplay between the superconducting gap and
normal-state pseudogap, the some main features of the doping and temperature dependence of the specific-heat, the
condensation energy, and the upper critical field are well reproduced. In particular, it is shown that in analogy to the
domelike shape of the doping dependence of the superconducting transition temperature, the maximal upper critical field
occurs around the optimal doping, and then decreases in both underdoped and overdoped regimes. Our results also show that
the humplike anomaly of the specific-heat near superconducting transition temperature in the underdoped regime can be
attributed to the emergence of the normal-state pseudogap in cuprate superconductors.
\end{abstract}


\maketitle

\section{Introduction}

The doping and temperature dependence of the thermodynamic properties for cuprate superconductors has been the subject of
much experimental and theoretical investigation \cite{Junod90}. In the conventional superconductors \cite{schrieffer83},
the absence of the low-energy electron excitations is reflected in the thermodynamic properties, such as the specific heat
$C_{\rm v}$. Although small deviations from exponential behavior have been observed in some conventional superconductors
at the low-temperatures, the specific heat of the most conventional superconductors is experimentally found to be
exponential at the low-temperatures, since the conventional superconductors are fully gaped \cite{schrieffer83}. However,
the characteristic feature of cuprate superconductors is the existence of four nodes on the Fermi surface
\cite{damascelli03}, where the d-wave superconducting (SC) gap vanishes
$\bar{\Delta}({\bf k})|_{\rm at~nodes}=\bar{\Delta}({\rm cos}k_{x}-{\rm cos}k_{y})|_{\rm at~ nodes}=0$. In this case, the
thermodynamic properties for cuprate superconductors are decreased as some power of the temperature. Moreover, since
cuprate superconductors are the doped Mott insulators, obtained by chemically adding charge carriers to a strongly
correlated antiferromagnetic insulating state \cite{damascelli03}, the thermodynamic properties of cuprate
superconductors mainly depend on the extent of dopings, and the regimes have been classified into the underdoped,
optimally doped, and overdoped, respectively.

Experimentally, by virtue of systematic studies using the heat capacity measurement technique, some essential features
of the evolution of the specific-heat in cuprate superconductors with doping and temperature have been established now
\cite{Loram94,Moler97,Junod99,Wen09,Loram00}, where the specific-heat in both the SC-state and normal-state in the
underdoped regime shows anomalous properties when compared with the case in the optimally doped and overdoped regimes.
The early heat capacity measurements \cite{Loram94,Moler97,Junod99} showed that the specific-heat of cuprate
superconductors in the underdoped regime was highly anomalous and deviated strongly from a simple d-wave
Bardeen-Cooper-Schrieffer (BCS) form, and the anomalies are a marked reduction in the size of the specific-heat jump
near the SC transition temperature $T_{c}$ and a depression in the normal state above $T_{c}$. Later, the heat capacity
measurements \cite{Wen09} indicated that the specific-heat has a humplike anomaly near $T_{c}$ and behaves as a long tail
in the underdoped regime, while in the heavily overdoped regime, the anomaly ends sharply just near $T_{c}$. Moreover,
it was argued these anomalous specific-heat results as evidence that in the underdoped regime the pseudogap is an
intrinsic feature of the normal-state density of states that compets with the SC condensate for the low energy spectral
weight \cite{Loram94,Loram00}. Furthermore, by virtue of the magnetization measurement technique, the value of the upper
critical field and its doping and temperature dependence have been observed for all the temperature $T\leq T_{c}$
throughout the SC dome \cite{Lee93,Niderost98,Nakagawa98,Luo00,Miura02,Wang06,Wen08}, where at the low temperatures, the
upper critical field becomes larger as one moves from the underdoped regime to the optimal doping, and then falls with
increasing doping in the overdoped regime, forming a domelike shape doping dependence like $T_{c}$. However, at a given
doping concentration, the temperature dependence of the upper critical field follows qualitatively the pair gap
temperature dependence \cite{Lee93,Niderost98,Nakagawa98,Luo00,Miura02,Wang06,Wen08}. Although the doping and temperature
dependence of the thermodynamic properties for cuprate superconductors are well-established experimentally \cite{Loram94,Moler97,Junod99,Wen09,Loram00,Lee93,Niderost98,Nakagawa98,Luo00,Miura02,Wang06,Wen08} and an agreement has
emerged theoretically that the specific-heat of cuprate superconductors in the underdoped regime is not describable
within the simple d-wave BCS formalism, its full understanding is still a challenging issue. In particular, the
specific-heat of cuprate superconductors has been calculated based on a phenomenological theory of the normal-state
pseudogap state \cite{Tesanovic08}, and the results show that the strong suppression of the specific-heat jump
near $T_{c}$ and the corresponding reduction in condensation energy with increased underdoping can be understood as due to
the emergence of a pseudogap. However, up to now, the thermodynamic properties of cuprate superconductors have not been
treated starting from a microscopic SC theory, and no explicit calculations of the doping and temperature dependence of
the upper critical field has been made so far.

In our recent study \cite{feng11}, the interplay between the SC gap and normal-state pseudogap in cuprate superconductors
is studied based on the kinetic energy driven SC mechanism \cite{feng0306}, where we show that the interaction between
charge carriers and spins directly from the kinetic energy by exchanging spin excitations in the higher power of the
doping concentration induces the normal-state pseudogap state in the particle-hole channel and the SC-state in the
particle-particle channel, then there is a coexistence of the SC gap and normal-state pseudogap in the whole SC dome. In
particular, this normal-state pseudogap is closely related to the quasiparticle coherent weight, and therefore it is a
necessary ingredient for superconductivity in cuprate superconductors. Moreover, both the normal-state pseudogap and SC
gap are dominated by one energy scale, and they are the result of the strong electron correlation. In this paper, we
start from this theoretical framework \cite{feng11}, and then provide a natural explanation to the doping and temperature
dependence of the thermodynamic properties in cuprate superconductors. We evaluate explicitly the specific-heat and upper
critical field, and qualitatively reproduced some main features of the heat capacity and magnetization measurements on
cuprate superconductors
\cite{Loram94,Moler97,Junod99,Wen09,Loram00,Lee93,Niderost98,Nakagawa98,Luo00,Miura02,Wang06,Wen08}. In particular, we
show that in analogy to the domelike shape of the doping dependence of the SC transition temperature, the upper critical
field increases with increasing doping in the underdoped regime, and reaches a maximum in the optimal doping, then
decreases with increasing doping in the overdoped regime.

The rest of this paper is organized as follows. We present the basic formalism in Section \ref{framework}, and then the
quantitative characteristics of the doping and temperature dependence of the thermodynamic properties are discussed in
Section \ref{thermodynamic}, where we show that the humplike anomaly of the specific-heat near $T_{c}$ in the underdoped
regime can be attributed to the emergence of the normal-state pseudogap in cuprate superconductors. Finally, we give a
summary in Section \ref{conclusions}.

\section{Theoretical framework}\label{framework}

In cuprate superconductors, the characteristic feature is the presence of the CuO$_{2}$ plane \cite{damascelli03}. In
this case, it is commonly accepted that the essential physics of the doped CuO$_{2}$ plane \cite{anderson87} is captured
by the $t$-$J$ model on a square lattice,
\begin{eqnarray}\label{tjham}
H&=&-t\sum_{l\hat{\eta}\sigma}C^{\dagger}_{l\sigma}C_{l+\hat{\eta}\sigma}+t'\sum_{l\hat{\eta}'\sigma}
C^{\dagger}_{l\sigma}C_{l+\hat{\eta}'\sigma}+\mu\sum_{l\sigma} C^{\dagger}_{l\sigma}C_{l\sigma}\nonumber\\
&+&J\sum_{l\hat{\eta}}{\bf S}_{l}\cdot {\bf S}_{l+\hat{\eta}},
\end{eqnarray}
where $\hat{\eta}=\pm\hat{x},\pm\hat{y}$, $\hat{\eta}'=\pm\hat{x}\pm\hat{y}$, $C^{\dagger}_{l\sigma}$ ($C_{l\sigma}$) is
the electron creation (annihilation) operator, ${\bf S}_{l}=(S^{x}_{l},S^{y}_{l}, S^{z}_{l})$ are spin operators, and
$\mu$ is the chemical potential. This $t$-$J$ model (\ref{tjham}) is in the Hilbert subspace with no doubly occupied
electron states on the same site, i.e., $\sum_{\sigma}C^{\dagger}_{l\sigma}C_{l\sigma}\leq 1$. To incorporate this
electron single occupancy local constraint, the charge-spin separation (CSS) fermion-spin theory \cite{feng0304,feng08}
has been proposed, where the physics of no double occupancy is taken into account by representing the electron as a
composite object created by $C_{l\uparrow}= h^{\dagger}_{l\uparrow}S^{-}_{l}$ and
$C_{l\downarrow}=h^{\dagger}_{l\downarrow}S^{+}_{l}$, with the spinful fermion operator
$h_{l\sigma}=e^{-i\Phi_{l\sigma}}h_{l}$ that describes the charge degree of freedom of the electron together with some
effects of spin configuration rearrangements due to the presence of the doped hole itself (charge carrier), while the
spin operator $S_{l}$ represents the spin degree of freedom of the electron, then the electron single occupancy local
constraint is satisfied in analytical calculations. In this CSS fermion-spin representation, the $t$-$J$ model
(\ref{tjham}) can be expressed as,
\begin{eqnarray}\label{cssham}
H&=&t\sum_{l\hat{\eta}}(h^{\dagger}_{l+\hat{\eta}\uparrow}h_{l\uparrow}S^{+}_{l}S^{-}_{l+\hat{\eta}}
+h^{\dagger}_{l+\hat{\eta}\downarrow}h_{l\downarrow}S^{-}_{l}S^{+}_{l+\hat{\eta}})\nonumber\\
&-&t'\sum_{l\hat{\eta}'}(h^{\dagger}_{l+\hat{\eta}'\uparrow}h_{l\uparrow}S^{+}_{l}S^{-}_{l+\hat{\eta}'}
+h^{\dagger}_{l+\hat{\eta}'\downarrow}h_{l\downarrow}S^{-}_{l}S^{+}_{l+\hat{\eta}'})\nonumber\\
&-&\mu\sum_{l\sigma} h^{\dagger}_{l\sigma}h_{l\sigma}+J_{{\rm eff}}\sum_{l\hat{\eta}}{\bf S}_{l}
\cdot {\bf S}_{l+\hat{\eta}},
\end{eqnarray}
where $J_{{\rm eff}}=(1-\delta)^{2}J$, and
$\delta=\langle h^{\dagger}_{l\sigma}h_{l\sigma}\rangle=\langle h^{\dagger}_{l}h_{l}\rangle$ is the charge carrier doping
concentration.

For discussions of the doping and temperature dependence of the thermodynamic properties in cuprate superconductors, we
need to evaluate the internal energy of the system, which can be separated into two parts in the CSS fermion-spin
representation as,
\begin{eqnarray}\label{energy}
U_{\rm total}(T,\delta)=U_{\rm charge}(T,\delta)+U_{\rm spin}(T,\delta),
\end{eqnarray}
with $U_{\rm charge}(T,\delta)$ and $U_{\rm spin}(T,\delta)$ are the corresponding contributions from the charge carriers
and spins, respectively, and can be expressed as,
\begin{subequations}\label{energy1}
\begin{eqnarray}
U_{\rm charge}(T,\delta)&=&2\int\limits_{-\infty}^\infty {{\rm{d}}\omega\over 2\pi}\omega
\rho_{\rm charge}(\omega,T,\delta)n_{F}(\omega), \\
U_{\rm spin}(T,\delta)&=&\int\limits_{-\infty}^\infty {{\rm{d}}\omega\over 2\pi}\omega\rho_{\rm spin}(\omega,T,\delta)
n_{B}(\omega),
\end{eqnarray}
\end{subequations}
where $\rho_{\rm charge}(\omega,T,\delta)$ is the charge carrier density of states, $\rho_{\rm spin}(\omega,T,\delta)$ is
the spin density of states, and the two in the charge carrier part of the internal energy is for spin degeneracy, while
$n_{F}(\omega)$ and $n_{B}(\omega)$ are the fermion and boson distribution functions, respectively.

As in the conventional superconductors, the key phenomenon occurring in cuprate superconductors in the SC state is the
pairing of charge carriers \cite{damascelli03}. The system of charge carriers forms pairs of bound charge carriers in the
SC state, while the pairing means that there is an attraction between charge carriers. For a microscopic description of
the SC-state of cuprate superconductors, the kinetic energy driven SC mechanism \cite{feng0306} has been developed based
on the CSS fermion-spin theory \cite{feng0304,feng08}, where the attraction between charge carriers mediated by the spin
excitations occurs directly through the kinetic energy, then the electron Cooper pairs originating from the charge
carrier pairing state are due to the charge-spin recombination, and their condensation reveals the SC ground-state. In
particular, the SC transition temperature is identical to the charge carrier pair transition temperature. Within this
kinetic energy driven SC mechanism, we have discussed the interplay between the SC-state and normal-state pseudogap state
in cuprate superconductors \cite{feng11}, and the obtained phase diagram with the two-gap feature is consistent
qualitatively with the experimental data observed on different families of cuprate superconductors \cite{Hufner08}.
Following these previous discussions \cite{feng11,feng0306}, the full charge carrier diagonal and off-diagonal Green's
functions and the mean-field (MF) spin Green's functions can be obtained explicitly as,
\begin{subequations}\label{green-functions}
\begin{eqnarray}
g({\bf k},\omega)&=&{U^{2}_{1{\rm h}{\bf k}}\over \omega-E_{1{\rm h}{\bf k}}}+{V^{2}_{1{\rm h}{\bf k}}\over\omega
+E_{1{\rm h}{\bf k}}}\nonumber\\
&+&{U^{2}_{2{\rm h}{\bf k}}\over\omega-E_{2{\rm h}{\bf k}}}+{V^{2}_{2{\rm h}{\bf k}}\over\omega+E_{2{\rm h}{\bf k}}}, \\
\Gamma^{\dagger}({\bf k},\omega)&=&-{\alpha_{1{\bf k}}\bar{\Delta}_{\rm h}({\bf k})\over 2E_{1{\rm h}{\bf k}}}
\left ({1\over \omega-E_{1{\rm h}{\bf k}}}-{1\over \omega+E_{1{\rm h}{\bf k}}}\right )\nonumber\\
&+&{\alpha_{2{\bf k}}\bar{\Delta}_{\rm h}({\bf k})\over 2 E_{2{\rm h}{\bf k}}}\left ({1\over \omega-E_{2{\rm h}{\bf k}}}
-{1\over \omega+E_{2{\rm h}{\bf k}}}\right ),~~~~~\\
D^{(0)}({\bf p},\omega)&=&{B({\bf p})\over 2\omega({\bf p})} \left ({1\over\omega-\omega({\bf p})}-{1\over\omega
+\omega({\bf p})}\right ),\\
D^{(0)}_{z}({\bf p},\omega)&=&{B_{z}({\bf p})\over 2\omega_{z}({\bf p})} \left ({1\over\omega-\omega_{z}({\bf p})}
-{1\over\omega+\omega_{z}({\bf p})}\right ),
\end{eqnarray}
\end{subequations}
respectively, where
$\alpha_{1{\bf k}}=(E^{2}_{1{\rm h}{\bf k}}-M^{2}_{\bf k})/(E^{2}_{1{\rm h}{\bf k}}-E^{2}_{2{\rm h}{\bf k}})$,
$\alpha_{2{\bf k}}=(E^{2}_{2{\rm h}{\bf k}}-M^{2}_{\bf k})/(E^{2}_{1{\rm h}{\bf k}}-E^{2}_{2{\rm h}{\bf k}})$, and there
are four coherent charge carrier quasiparticle bands due to the presence of the normal-state pseudogap and SC gap,
$E_{1{\rm h}{\bf k}}$, $-E_{1{\rm h}{\bf k}}$, $E_{2{\rm h}{\bf k}}$, and $-E_{2{\rm h}{\bf k}}$, with
$E_{1{\rm h}{\bf k}}=\sqrt{[\Omega_{\bf k}+\Theta_{\bf k}]/2}$,
$E_{2{\rm h}{\bf k}}=\sqrt{[\Omega_{\bf k}-\Theta_{\bf k}]/2}$, and the kernel functions,
\begin{subequations}
\begin{eqnarray}
\Omega_{\bf k}&=&\xi^{2}_{\bf k}+M^{2}_{\bf k}+8\bar{\Delta}^{2}_{\rm pg}({\bf k})+\bar{\Delta}^{2}_{\rm h}({\bf k}),\\
\Theta_{\bf k}&=&\sqrt{(\xi^{2}_{\bf k}-M^{2}_{\bf k})\beta_{1{\bf k}}+16\bar{\Delta}^{2}_{\rm pg}({\bf k})\beta_{2{\bf k}}
+\bar{\Delta}^{4}_{\rm h}({\bf k})},~~~~~~
\end{eqnarray}
\end{subequations}
where $\beta_{1{\bf k}}=\xi^{2}_{\bf k}-M^{2}_{\bf k}+2\bar{\Delta}^{2}_{\rm h}({\bf k})$,
$\beta_{2{\bf k}}=(\xi_{\bf k}-M_{\bf k})^{2}+\bar{\Delta}^{2}_{\rm h}({\bf k})$, the MF charge carrier excitation
spectrum $\xi_{{\bf k}}=Zt\chi_{1}\gamma_{{\bf k}}-Zt'\chi_{2}\gamma_{{\bf k}}'-\mu$, the spin correlation functions
$\chi_{1}=\langle S_{i}^{+}S_{i+\hat{\eta}}^{-}\rangle$, $\chi_{2}=\langle S_{i}^{+}S_{i+\hat{\eta}'}^{-}\rangle$,
$\gamma_{{\bf k}}=(1/Z)\sum_{\hat{\eta}}e^{i{\bf k}\cdot \hat{\eta}}$,
$\gamma_{{\bf k}}'= (1/Z)\sum_{\hat{\eta}'}e^{i{\bf k} \cdot\hat{\eta}'}$, $Z$ is the number of the nearest neighbor or
second-nearest neighbor sites, the effective charge carrier d-wave pair gap
$\bar{\Delta}_{\rm h}({\bf k})=\bar{\Delta}_{\rm h}({\rm cos} k_{x}-{\rm cos}k_{y})/2$, and the effective normal-state
pseudogap $\bar{\Delta}_{\rm pg}({\bf k})$ and energy spectrum $M_{\bf k}$ have been given in Ref. \onlinecite{feng11},
while the coherence factors,
\begin{subequations}\label{coherence-factors}
\begin{eqnarray}
U^{2}_{1{\rm h}{\bf k}}&=&{1\over 2}\left [\alpha_{1{\bf k}}\left (1+{\xi_{\bf k}\over E_{1{\rm h}{\bf k}}}\right )-
\alpha_{3{\bf k}}\left (1+{M_{\bf k}\over E_{1{\rm h}{\bf k}}}\right )\right ],\\
V^{2}_{1{\rm h}{\bf k}}&=&{1\over 2}\left [\alpha_{1{\bf k}}\left (1-{\xi_{\bf k}\over E_{1{\rm h}{\bf k}}}\right )-
\alpha_{3{\bf k}}\left (1-{M_{\bf k}\over E_{1{\rm h}{\bf k}}}\right )\right ],\\
U^{2}_{2{\rm h}{\bf k}}&=&-{1\over 2}\left [\alpha_{2{\bf k}}\left (1+{\xi_{\bf k}\over E_{2{\rm h}{\bf k}}}\right )-
\alpha_{3{\bf k}}\left (1+{M_{\bf k}\over E_{2{\rm h}{\bf k}}}\right )\right ],\\
V^{2}_{2{\rm h}{\bf k}}&=&-{1\over 2}\left [\alpha_{2{\bf k}}\left (1-{\xi_{\bf k}\over E_{2{\rm h}{\bf k}}}\right )-
\alpha_{3{\bf k}}\left (1-{M_{\bf k}\over E_{2{\rm h}{\bf k}}}\right )\right ],~~~~~
\end{eqnarray}
\end{subequations}
satisfy the sum rule: $U^{2}_{1{\rm h}{\bf k}}+V^{2}_{1{\rm h}{\bf k}}+U^{2}_{2{\rm h}{\bf k}}+V^{2}_{2{\rm h}{\bf k}}=1$,
where $\alpha_{3{\bf k}}=[2\bar{\Delta}_{\rm pg}({\bf k})]^{2}/(E^{2}_{1{\rm h}{\bf k}}-E^{2}_{2{\rm h}{\bf k}})$, while
the functions $B({\bf p})$ and $B_{z}({\bf p})$ and the MF spin excitations $\omega({\bf p})$ and $\omega_{z}({\bf p})$
have been given in Refs. \onlinecite{guo07} and \onlinecite{feng08}, then all order parameters and chemical potential are
determined by the self-consistent calculation.

With the helps of these full charge carrier diagonal Green's function and spin Green's functions in Eq.
(\ref{green-functions}), the charge carrier density of states and spin density of states can be obtained as,
\begin{subequations}\label{density}
\begin{eqnarray}
\rho_{\rm charge}(\omega,T,\delta)&=&{1\over N}\sum_{{\bf k}}A_{\rm charge}({\bf k},\omega,T,\delta), \\
\rho_{\rm spin}(\omega,T,\delta)&=&{1\over 2N}\sum_{{\bf k}}[A_{\rm spin}({\bf k},\omega,T,\delta)\nonumber\\
&+&A^{(z)}_{\rm spin}({\bf k},\omega,T,\delta)],
\end{eqnarray}
\end{subequations}
respectively, where the charge carrier spectral function
$A_{\rm charge}({\bf k},\omega,T,\delta)=-2{\rm Im}g({\bf k},\omega)$, and the spin spectral functions
$A_{\rm spin}({\bf k},\omega,T,\delta)=-2{\rm Im}D^{(0)}({\bf k},\omega)$ and
$A^{(z)}_{\rm spin}({\bf k},\omega,T,\delta)=-2{\rm Im}D^{(0)}_{z}({\bf k},\omega)$. Substituting these corresponding
charge carrier density of states and the spin density of states into Eq. (\ref{energy1}), and then incorporating the
self-consistent equations \cite{guo07,feng11}, we obtain the internal energy of cuprate superconductors in the
SC-state as,
\begin{eqnarray}\label{senergy}
U^{(s)}_{\rm total}(T,\delta)&=&-{1\over N}\sum_{\bf k}[E_{1{\rm h}{\bf k}}(\alpha_{1{\bf k}}-\alpha_{3{\bf k}}){\rm th}
({1\over 2}\beta E_{1{\rm h}{\bf k}})\nonumber\\
&-&E_{2{\rm h}{\bf k}}(\alpha_{2{\bf k}}-\alpha_{3{\bf k}}){\rm th}({1\over 2}\beta
E_{2{\rm h}{\bf k}})]\nonumber\\
&+&{1\over N}\sum_{\bf k}\xi_{{\bf k}}+ZJ_{{\rm eff}}(\chi_{1}+\chi^{z}_{1}),
\end{eqnarray}
with the spin correlation function $\chi^{z}_{1}=\langle S^{z}_{i}S^{z}_{i+\hat{\eta}}\rangle$. In the normal-state,
where the charge carrier pair gap $\bar{\Delta}_{\rm h}=0$, this internal energy is reduced as,
\begin{eqnarray}\label{nenergy}
U^{(n)}_{\rm total}(T,\delta)&=&-{1\over N}\sum_{\bf k}[E^{+}_{{\rm h}{\bf k}}\alpha^{(n)}_{1{\bf k}}{\rm th}({1\over 2}
\beta E^{+}_{{\rm h}{\bf k}})\nonumber\\
&-&E^{-}_{{\rm h}{\bf k}}\alpha^{(n)}_{2{\bf k}}{\rm th}({1\over 2}
\beta E^{-}_{{\rm h}{\bf k}})]\nonumber\\
&+&{1\over N}\sum_{\bf k}\xi_{{\bf k}}+ZJ_{{\rm eff}}(\chi_{1}+\chi^{z}_{1}),
\end{eqnarray}
with $\alpha^{(n)}_{1{\bf k}}=(E^{+}_{{\rm h}{\bf k}}+M_{\bf k})/(E^{+}_{{\rm h}{\bf k}}-E^{-}_{{\rm h}{\bf k}})$,
$\alpha^{(n)}_{2{\bf k}}=(E^{-}_{{\rm h}{\bf k}}+M_{\bf k})/(E^{+}_{{\rm h}{\bf k}}-E^{-}_{{\rm h}{\bf k}})$,
$E^{+}_{{\rm h}{\bf k}}=[\xi_{{\bf k}}-M_{\bf k}+\sqrt{(\xi_{{\bf k}}+M_{\bf k})^{2}+16\bar{\Delta}^{2}_{\rm pg}({\bf k})}]/2$,
and
$E^{-}_{{\rm h}{\bf k}}=[\xi_{{\bf k}}-M_{\bf k}-\sqrt{(\xi_{{\bf k}}+M_{\bf k})^{2}+16\bar{\Delta}^{2}_{\rm pg}({\bf k})}]/2$.

\section{Quantitative characteristics of the thermodynamic properties}\label{thermodynamic}

\begin{figure}[h!]
\includegraphics[scale=0.45]{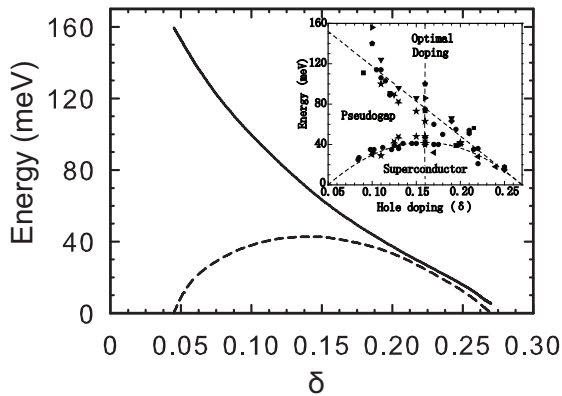}
\caption{The magnitude of the effective normal-state pseudogap parameter (2$\bar{\Delta}_{\rm pg}$) (solid line) and
effective charge carrier pair gap parameter ($2\bar{\Delta}_{\rm h}$) (dashed line) as a function of doping for
temperature $T=0.002J$ with parameters $t/J=2.5$, $t'/t=0.3$, and $J=110$meV. Inset: the corresponding experimental
data of cuprate superconductors taken from Ref. \onlinecite{Hufner08}.
\label{fig1}}
\end{figure}

\begin{figure}[h!]
\includegraphics[scale=0.45]{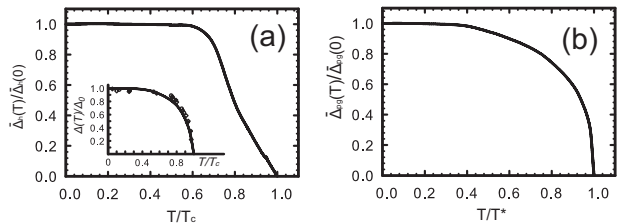}
\caption{(a) The effective charge carrier pair gap parameter and (b) the effective normal-state pseudogap parameter as a
function of temperature in the doping concentration $\delta=0.09$ with $t/J=2.5$, $t'/t=0.3$, and $J=110$meV. Inset in
(a): the corresponding experimental data of the pair gap parameter for the underdoped
Bi$_{2}$Sr$_{2}$Ca$_{2}$Cu$_{3}$O$_{10+\delta}$ taken from Ref. \onlinecite{Vishik10}.
\label{fig2}}
\end{figure}

In this section, we discuss some basic behaviors of the doping and temperature dependence of the thermodynamic properties
in cuprate superconductors. In cuprate superconductors, although the values of $J$, $t$, and $t'$ are believed to vary
somewhat from compound to compound \cite{damascelli03}, however, as in our previous studies \cite{feng11}, the commonly
used parameters in this paper are chosen as $t/J=2.5$, $t'/t=0.3$, and $J=110$meV for a qualitative discussion. In this
case, for a complement of the previous analysis of the interplay between the SC gap and normal-state pseudogap in cuprate
superconductors \cite{feng11}, we replot the magnitude of the effective normal-state pseudogap parameter
(2$\bar{\Delta}_{\rm pg}$) (solid line) and effective charge carrier pair gap parameter ($2\bar{\Delta}_{\rm h}$) (dashed
line) as a function of doping for temperature $T=0.002J$ in Fig. \ref{fig1} in comparison with the corresponding
experimental results \cite{Hufner08} observed on different families of cuprate superconductors (inset). In cuprate
superconductors, the charge carrier pairing gap parameter measures the strength of the binding of charge carriers into
the charge carrier pairs, while the normal-state pseudogap is closely related to the unusual physical properties. Our
theoretical results in Fig. \ref{fig1} reproduce qualitatively the two-gap feature observed on cuprate superconductors
\cite{Hufner08}, and show that the effective charge carrier pair gap parameter increases with increasing doping in the
underdoped regime, and reaches a maximum in the optimal doping, then decreases with increasing doping in the overdoped
regime \cite{feng0306,guo07}. However, in contrast to the case of the effective charge carrier pair gap parameter in the
underdoped regime, the magnitude of the effective normal-state pseudogap parameter smoothly increases with decreasing
doping in the underdoped regime, this leads to that the magnitude of the effective normal-state pseudogap parameter is
much larger than the effective charge carrier pair gap parameter in the underdoped regime. Moreover, the magnitude of the
normal-state pseudogap parameter seems to merge with the charge carrier pair gap parameter in the overdoped regime,
eventually disappearing together with superconductivity at the doping concentrations larger than the doping concentration
$\delta\sim 0.27$. Furthermore, these effective charge carrier pair gap parameter and effective normal-state pseudogap
parameter are strongly temperature dependent. To show this point clearly, we plot (a) the effective charge carrier pair
gap parameter and (b) the effective normal-state pseudogap parameter as a function of temperature at the doping
concentration $\delta=0.09$ in Fig. \ref{fig2}. For comparison, the corresponding experimental result of the pair gap
parameter \cite{Vishik10} for the underdoped Bi$_{2}$Sr$_{2}$Ca$_{2}$Cu$_{3}$O$_{10+\delta}$ is also shown in Fig.
\ref{fig2} [inset in (a)]. Obviously, both the effective charge carrier pair gap parameter and the effective normal-state
pseudogap parameter have a similar temperature dependence, and they decreases with increasing temperatures, however, the
effective charge carrier pair gap parameter vanishes at $T_{c}$, while the effective normal-state pseudogap parameter
vanishes at the normal-state pseudogap crossover temperature $T^{*}$, where $T^{*}$ is much larger than $T_{c}$ in the
underdoped regime \cite{feng11}.

\begin{figure}[h!]
\includegraphics[scale=0.45]{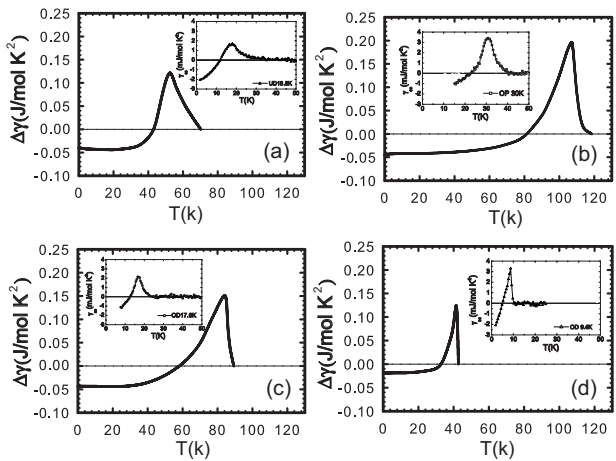}
\caption{The specific-heat coefficient as a function of temperature at (a) $\delta=0.09$, (b) $\delta=0.15$, (c)
$\delta=0.18$, and (d) $\delta=0.25$ with $t/J=2.5$, $t'/t=0.3$, and $J=110$meV. Insets: the corresponding experimental
data of Bi$_{2}$Sr$_{2-x}$La$_{x}$CuO$_{6+\delta}$ taken from Ref. \onlinecite{Wen09}.
\label{fig3}}
\end{figure}

\begin{figure}[h!]
\includegraphics[scale=0.45]{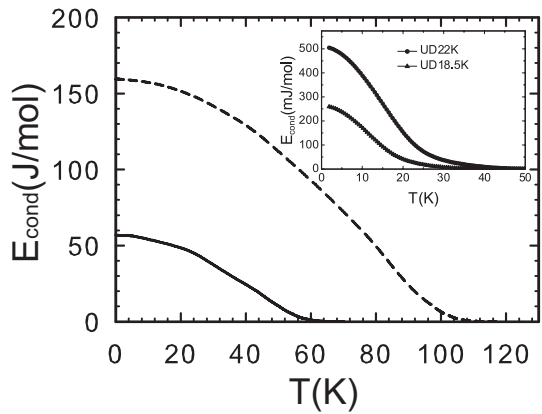}
\caption{The condensation energy as a function of temperature at $\delta=0.09$ (solid line) and $\delta=0.15$ (dashed
line) with $t/J=2.5$, $t'/t=0.3$, and $J=110$meV. Insets: the corresponding experimental data
for Bi$_{2}$Sr$_{2-x}$La$_{x}$CuO$_{6+\delta}$ in the underdoped regime taken from Ref. \onlinecite{Wen09}.
\label{fig4}}
\end{figure}

\subsection{Doping and temperature dependence of the specific-heat}

In the following discussions, we discuss the doping dependence of the specific-heat in cuprate superconductors. With
the helps of Eqs. (\ref{senergy}) and (\ref{nenergy}), the specific-heat can be obtained by evaluating the
temperature-derivative of the internal energies as,
\begin{eqnarray}\label{heat}
C^{(a)}_{\rm v}(T,\delta)={{\rm d}U^{(a)}(T,\delta)\over {\rm d}T}=\gamma_{a}(T,\delta)T,
\end{eqnarray}
where $a=s$, $n$, and $\gamma_{a}(T,\delta)$ is the doping and temperature dependence of the specific-heat coefficient.
In this case, we have performed a calculation for the specific-heat coefficient, and the results of
$\Delta\gamma(T,\delta)=\gamma_{s}(T,\delta)-\gamma_{n}(T,\delta)$ as a function of temperature in the underdoping (a)
$\delta=0.09$, (b) the optimal doping $\delta=0.15$, (c) the overdoping $\delta=0.18$, and (d) the heavily overdoping
$\delta=0.25$ are plotted in Fig. \ref{fig3} in comparison with the corresponding experimental data \cite{Wen09} for
Bi$_{2}$Sr$_{2-x}$La$_{x}$CuO$_{6+\delta}$ (inset). It is shown clearly that our present theoretical results capture
all essential qualitative features of the doping dependence of the specific-heat observed experimentally on cuprate
superconductors \cite{Loram94,Moler97,Junod99,Wen09,Loram00}. In the underdoped regime, the specific-heat jump near
$T_{c}$ is strongly suppressed, therefore there is no steplike specific-heat anomaly near $T_{c}$, instead, it shows a
humplike peak and remains as long tail of $\gamma_{s}(T,\delta)$. However, in the optimal doping, although the
specific-heat anomaly is still not a sharp steplike, it shows a symmetric peak, and therefore there is a tendency
towards to the steplike specific-heat anomaly with increasing doping. This tendency is particularly obvious in the
overdoped regime, where the long tail appeared in the underdoped regime becomes much shorter, then the specific-heat
anomaly ends near $T_{c}$ in the heavily overdoped regime, and a steplike BCS transition with the absence of the long
tail appears.

\subsection{Doping and temperature dependence of the condensation energy}

For a superconductor, it undergoes a transition from the normal-state to the SC-state because this transition can lower
the total free energy, and then the energy difference between the normal-state $F^{(n)}(T,\delta)$, extrapolated to zero
temperature, and the SC-state $F^{(s)}(T,\delta)$, is defined as the condensation energy $E_{\rm cond}(T,\delta)$,
\begin{eqnarray}\label{condensationenergy}
E_{\rm cond}(T,\delta)=F^{(n)}(T,\delta)-F^{(s)}(T,\delta),
\end{eqnarray}
where the free energies are obtained in terms of the corresponding internal energies in Eqs. (\ref{senergy}) and
(\ref{nenergy}) as,
\begin{eqnarray}\label{fenergy}
F^{(a)}(T,\delta)=U^{(a)}(T,\delta)-TS^{(a)}(T,\delta),
\end{eqnarray}
with the related entropy of the system is evaluated from the specific-heat coefficient in Eq. (\ref{heat}) as,
\begin{eqnarray}\label{entropy}
S^{(a)}(T,\delta)=\int\limits_{0}^{T}\gamma_{a}(T',\delta){\rm d}T'.
\end{eqnarray}
Alternatively, this condensation energy can also be obtained by integrating the difference in the specific-heat
coefficients in Eq. (\ref{heat}) in the normal-state and the SC-state \cite{Wen09} from zero temperature to $T_{c}$. In
this cae, we plot the condensation energy $E_{\rm cond}(T,\delta)$ as a function of temperature at $\delta=0.09$ (solid
line) and $\delta=0.15$ (dashed line) in Fig. \ref{fig4} in comparison with the corresponding experimental data
\cite{Wen09} for Bi$_{2}$Sr$_{2-x}$La$_{x}$CuO$_{6+\delta}$ in the underdoped regime (inset). Our results show that in
the underdoped regime, the condensation energy increases with increasing doping, then it follows qualitatively a pair
gap type temperature dependence, and disappears at $T_{c}$, in qualitative agreement with experimental data
\cite{Loram94,Moler97,Junod99,Wen09,Loram00}.

\subsection{Doping and temperature dependence of the upper critical field}\label{upperfield}

\begin{figure}[h!]
\includegraphics[scale=0.5]{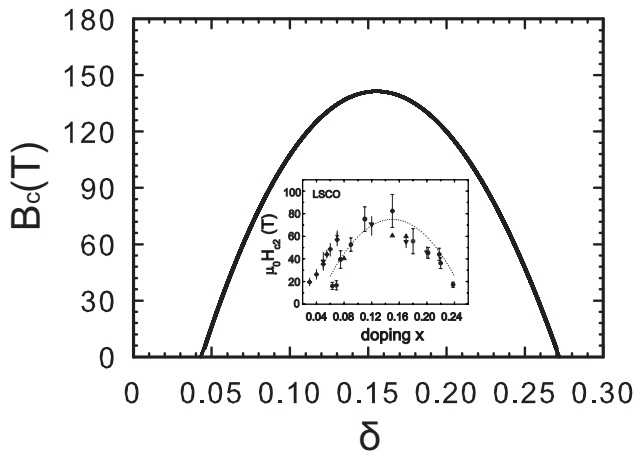}
\caption{Doping dependence of the upper critical field with $T=0.002J$ for $t/J=2.5$, $t'/t=0.3$, and $J=110$meV. Insets:
the experimental data for La$_{2-x}$Sr$_{x}$CuO$_{4}$ taken from Ref. \onlinecite{Wen08}.
\label{fig5}}
\end{figure}

\begin{figure}[h!]
\includegraphics[scale=0.5]{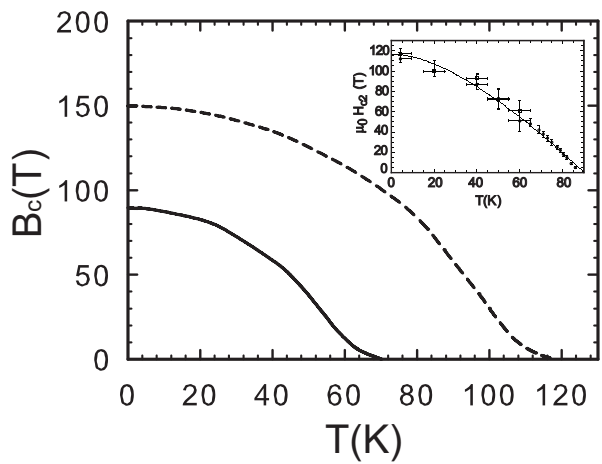}
\caption{Temperature dependence of the upper critical field at $\delta=0.09$ (solid line) and $\delta=0.15$ (dashed line)
for $t/J=2.5$, $t'/t=0.3$, and $J=110$meV. Inset: the corresponding experimental result for the slightly underdoped
YBa$_{2}$Cu$_{3}$O$_{7-\delta}$ taken from Ref. \onlinecite{Miura02}.
\label{fig6}}
\end{figure}

Now we turn to discuss the doping and temperature dependence of the upper critical field $B_{c}(T,\delta)$. For a given
doping concentration, the upper critical field is defined as the critical field that destroys the SC-state at the zero
temperature, therefore the upper critical field also measures the strength of the binding of charge carriers into charge
carrier pairs like the pair gap parameter. This doping and temperature dependence of the upper critical field is closely
related to the doping and temperature dependence of the condensation energy (\ref{condensationenergy}), and can be
obtained as,
\begin{eqnarray}\label{field}
{1\over 2\mu_{0}}B^{2}_{c}(T,\delta)=E_{\rm cond}(T,\delta).
\end{eqnarray}
In this case, we have performed firstly a calculation for the doping dependence of the upper critical field
$B_{c}(T,\delta)$ at the low temperatures, and the result of $B_{c}(T,\delta)$ as a function of doping with temperature
$T=0.002J$ is plotted in Fig. \ref{fig5}. For comparison, the corresponding experimental result \cite{Wen08} for
La$_{2-x}$Sr$_{x}$CuO$_{4}$ is also shown in Fig. \ref{fig5} (inset). Obviously, in analogy to the domelike shape of the
doping dependence of $T_{c}$ and pair gap parameter, the upper critical field increases with increasing doping in the
underdoped regime, and reaches a maximum in the optimal doping, then decreases with increasing doping in the overdoped
regime. This domelike shape of the doping dependence of the upper critical field is well consistent with the experimental
data \cite{Lee93,Niderost98,Nakagawa98,Luo00,Miura02,Wang06,Wen08}. Furthermore, we have discussed the temperature
dependence of the upper critical field, and the results of $B_{c}(T,\delta)$ as a function of temperature at $\delta=0.09$
(solid line) and $\delta=0.15$ (dashed line) are plotted in Fig. \ref{fig6} in comparison with the corresponding
experimental result \cite{Miura02} of the slightly underdoped YBa$_{2}$Cu$_{3}$O$_{7-\delta}$ (inset). Our results
indicate that as in the case of the temperature dependence of the condensation energy shown in Fig. \ref{fig4}, the upper
critical field $B_{c}(T)$ also follows qualitatively the pair gap type temperature dependence, i.e., it decreases with
increasing temperature, and vanishes at $T_{c}$, which is also qualitatively consistent with the experimental results
\cite{Lee93,Niderost98,Nakagawa98,Luo00,Miura02,Wang06,Wen08}. Since the upper critical field $B_{c}(T,\delta)$ (then
the condensation energy) is closely related to the difference between the free energies in the SC-state and normal-state,
the charge carrier pair gap parameter is relevant as shown in Eqs. (\ref{field}) and (\ref{fenergy}), i.e., the variation
of the upper critical field (then the condensation energy) with doping and temperature is coupled to the doping and
temperature dependence of the charge carrier pair gap parameter $\bar{\Delta}_{\rm h}$ in cuprate superconductors. In this
case, our present results of the upper critical field and its domelike shape of the doping dependence and pair gap type
temperature dependence also are a natural consequence of the results for the charge carrier pair gap parameter and its
domelike shape of the doping dependence and similar BCS type temperature dependence in the framework of the kinetic energy
driven SC mechanism \cite{feng0306} as shown in Fig. \ref{fig1} and Fig. \ref{fig2}.

\begin{figure}[h!]
\includegraphics[scale=0.5]{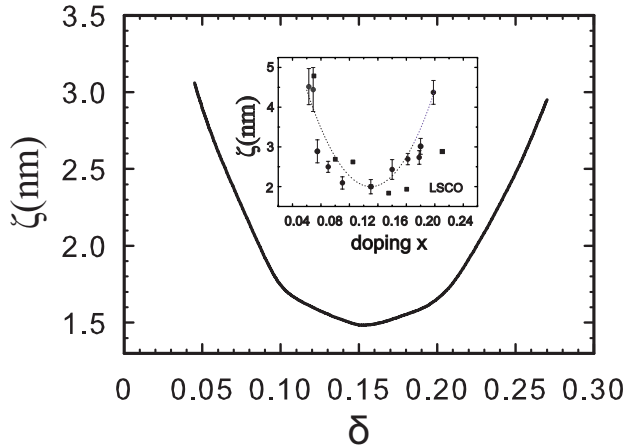}
\caption{The coherence length as a function of doping in $T=0.002J$ for $t/J=2.5$, $t'/t=0.3$, and $J=110$meV. Inset: the
experimental results for La$_{2-x}$Sr$_{x}$CuO$_{4}$ taken from Ref. \onlinecite{Wen08}.
\label{fig7}}
\end{figure}

The doping and temperature dependence of the coherence length $\zeta(T,\delta)$ is one of the important characteristic
parameters of cuprate superconductors. Although it can not be measured directly, it is closely related to the doping and
temperature dependence of the upper critical field as $\zeta^{2}(T,\delta)=\Phi_{0}/2\pi B_{c}(T,\delta)$, where
$\Phi_{0}=hc/2e$ is the magnetic flux quantum. In this case, with the help of the doping and temperature dependence of
the upper critical field in Eq. (\ref{field}), we can obtain the doping and temperature dependence of the coherence
length $\zeta(T,\delta)$, and the results of $\zeta(T,\delta)$ as a function of doping with temperature $T=0.002J$ are
plotted in Fig. \ref{fig7} in comparison with the corresponding experimental data \cite{Wen08} of
La$_{2-x}$Sr$_{x}$CuO$_{4}$ (inset). Obviously, the main feature of the doping dependence of the coherence length
$\zeta(T,\delta)$ obtained from the experiments \cite{Wen08} is reproduced, where in contrast to the case of the doping
dependence of the upper critical field, the coherence length $\zeta(T,\delta)$ in cuprate superconductors reaches a
minimum around the optimal doping, then grows in both the underdoped and overdoped regimes.

The essential physics of the humplike anomaly of the specific-heat near $T_{c}$ in cuprate superconductors in the
underdoped regime can be attributed to the emergence of the normal-state pseudogap \cite{feng11}. This follows a fact
that in the framework of the kinetic energy driven SC mechanism \cite{feng0306}, the normal-state pseudogap state is
particularly obvious in the underdoped regime as shown in Fig. \ref{fig1}, i.e., the magnitude of the normal-state
pseudogap is much larger than that of the pair gap in the underdoped regime, then it smoothly decreases upon increasing
doping as mentioned above. In this case, the sharp peak in the charge carrier density of states in the absence of the
normal-state pseudogap is spread out due to the band split in the presence of the normal-state pseudogap, reflecting a
suppression for the strength of the charge carrier density of states. In particular, this suppression for the strength
of the charge carrier density of states follows the same doping dependent behavior of the normal-state pseudogap, i.e.,
it decreases with increasing doping. This strong suppression for the strength of the charge carrier density of states
in the underdoped regime leads to a strong suppression of the specific-heat jump near $T_{c}$, then the humplike anomaly
near $T_{c}$ with a long tail in the underdoped regime is a natural consequence of the spread of the charge carrier
density of states. However, the range of this long tail decreases with increasing doping as the suppression for the
strength of the charge carrier density of states decreases upon increasing doping. In particular, in the heavily
overdoped regime, the normal-state pseudogap merges with the charge carrier pair gap parameter as shown in Fig.
\ref{fig1}. This reflects a fact that in the heavily overdoped regime, when the temperature $T=T_{c}$, the charge
carrier pair gap parameter $\bar{\Delta}_{\rm h}=0$, and at the same time, the normal-state pseudogap is negligible,
i.e., $\bar{\Delta}_{\rm pg}\approx 0$, this leads to a disappearance of the suppression for the strength of the charge
carrier density of states near $T_{c}$. In this case, the full charge carrier diagonal and off-diagonal Green's functions
(\ref{green-functions}) near $T_{c}$ can be induced as a simple d-wave BCS formalism \cite{feng0306,guo07},
\begin{subequations}\label{BCSform}
\begin{eqnarray}
g({\bf k},\omega)&=&{U^{2}_{{\rm h}{\bf k}}\over\omega-E_{{\rm h}{\bf k}}}+{V^{2}_{{\rm h}{\bf k}}\over\omega+
E_{{\rm h}{\bf k}}}, \\
\Gamma^{\dagger}({\bf k},\omega)&=&-{\bar{\Delta}_{\rm h}({\bf k})\over 2E_{{\rm h}{\bf k}}}\left ( {1\over\omega-
E_{{\rm h}{\bf k}}}-{1\over\omega + E_{{\rm h}{\bf k}}}\right ),~~~~~
\end{eqnarray}
\end{subequations}
although the pairing mechanism is driven by the kinetic energy by exchanging spin excitations, where the charge carrier
qasiparticle coherence factors $U^{2}_{{\rm h}{\bf k}}=(1+\xi_{{\bf k}}/E_{{\rm h}{\bf k}})/2$ and
$V^{2}_{{\rm h}{\bf k}}=(1-\xi_{{\bf k}}/E_{{\rm h}{\bf k}})/2$, and the charge carrier quasiparticle spectrum
$E_{{\rm h}{\bf k}}=\sqrt{\xi^{2}_{{\bf k}}+\mid\bar{\Delta}_{\rm h}({\bf k})\mid^{2}}$. This simple d-wave BCS formalism
(\ref{BCSform}) leads to that the specific-heat anomaly ends sharply just near $T_{c}$. This is also why the humplike
anomaly near $T_{c}$ with a long tail of the specific-heat appeared obviously in the underdoped regime is absent
in the heavily overdoped regime.

\section{Conclusions}\label{conclusions}

Based on the $t$-$J$ model, we have discussed the doping and temperature dependence of the thermodynamic properties in
cuprate superconductors. By considering the interplay between the SC gap and normal-state pseudogap within the framework
of the kinetic energy driven SC mechanism, we have reproduced qualitatively some main features of the doping and
temperature dependence of the specific-heat, the condensation energy, and the upper critical field. The specific-heat
shows a humplike peak and remains as long tail in the underdoped regime, however, this long tail is absent in the heavily
overdoped regime, and then the specific-heat shows a steplike BCS transition, while the condensation energy increases with
increasing doping in the underdoped regime, and follows a pair gap type temperature dependence. Moreover, in analogy to
the domelike shape of the doping dependence of $T_{c}$, the maximal upper critical field occurs around the optimal doping,
and then decreases in both underdoped and overdoped regimes. Our results also show that the striking behavior of the
specific-heat humplike anomaly near $T_{c}$ is closely related to the doping and temperature dependence of the
normal-state pseudogap. Since the knowledge of the doping and temperature dependence of the thermodynamic properties in
cuprate superconductors is of considerable importance as a test for theories of the normal-state and SC-state, the
qualitative agreement between the present theoretical results and experimental data also provides an important
confirmation of the nature of the SC phase of cuprate superconductors as a coexistence of the d-wave SC-state and
normal-state pseudogap state in the whole SC dome within the kinetic energy driven SC mechanism.

\acknowledgments

The authors would like to thank Dr. Zheyu Huang for helpful discussions. This work was supported by the funds from the
Ministry of Science and Technology of China under Grant Nos. 2011CB921700 and 2012CB821403, and the National Natural
Science Foundation of China under Grant No. 11074023.

\end{document}